\documentclass[12pt,preprint]{aastex}

\newcommand{\msun}{$\mathrm{\,M_\odot}$}

\newcommand{\rcb}{$\rho$~CrB}
\newcommand{\kps}{$\mathrm{\,km\,s^{-1}}$}
\newcommand{\mps}{$\mathrm{\,m\,s^{-1}}$}
\newcommand{\um}{$\mathrm{\,\mu{}m}$}

\shorttitle{$\rho$ CrB Companion Flux Ratio}
\shortauthors{Bender et al.}

\begin{document}

\title{An Upper Bound on the Flux Ratio of \rcb's Companion at 1.6\um}

\slugcomment{Accepted for publication in the Astronomical Journal}

\author{C. Bender\altaffilmark{1}, M. Simon\altaffilmark{1},
L. Prato\altaffilmark{2,3}, T. Mazeh\altaffilmark{4}, and
S. Zucker\altaffilmark{5,6}}

\altaffiltext{1}{Department of Physics and Astronomy, State University
of New York at Stony Brook, Stony Brook, NY 11794-3800;
cbender@mail.astro.sunysb.edu, michal.simon@sunysb.edu.}

\altaffiltext{2}{Department of Physics and Astronomy, University of
California at Los Angeles, Los Angeles, CA 90095-1562.}

\altaffiltext{3}{Current address: Lowell Observatory, 1400 W. Mars
Hill Rd., Flagstaff, AZ 86001; lprato@lowell.edu.}

\altaffiltext{4}{Department of Physics and Astronomy, Tel Aviv
University, Tel Aviv, Israel 69978; mazeh@wise1.tau.ac.il.}

\altaffiltext{5}{Department of Geophysics and Planetary Sciences, Tel
Aviv University, Tel Aviv, Israel 69978.}

\altaffiltext{6}{Current address: Geneva Observatory, CH-1290 Sauverny, 
Switzerland; Shay.Zucker@obs.unige.ch.}

\begin{abstract}

We use high resolution infrared spectroscopy to investigate the 2001
report by Gatewood and colleagues that \rcb{}'s candidate extrasolar
planet companion is really a low-mass star with mass
$0.14\pm0.05$\msun.  We do not detect evidence of such a companion;
the upper bounds on the (companion/primary) flux ratio at 1.6\um{} are
less than 0.0024 and 0.005 at the 90 and 99\% confidence levels,
respectively.  Using the $H$-band mass-luminosity relationship
calculated by Baraffe and colleagues, the corresponding upper limits
on the companion mass are 0.11 and 0.15\msun.  Our results indicate
that the infrared spectroscopic technique can detect companions in
binaries with flux ratios as low as 0.01 to 0.02.

\end{abstract}

\keywords{binaries: spectroscopic --- stars: individual ($\rho$ Coronae Borealis) --- techniques: radial velocities, spectroscopic}

\section{Introduction\label{intro}}

$\rho$ Coronae Borealis (\rcb, HIP 78459), a well-studied G0V star
with an estimated mass of 1.0\msun{} and an age of 10\,Gyr, was among
the first stars identified with an extrasolar planet (ESP) candidate
\citep[hereafter N97]{noyes97}.  If the mass of the companion, $m$, is
small relative to \rcb, the system's mass function indicates that
$m~sin~i= 1.1\,M_{Jupiter}$, where $i$ is the orbital inclination.
For a random distribution of orbital plane inclinations, the
expectation value of $m$ is $(4/\pi) \times 1.1\,M_{Jupiter}$,
justifying the interpretation that \rcb's companion is an ESP.

Subsequently, \citet[hereafter GHB]{gatewood01} used data from {\it
Hipparcos} and the ground-based Multichannel Astrometric Photometer
\citep{G87} to measure the astrometric signature of the reflex motion
of \rcb.  They derived an astrometric orbit with a semi-major axis of
$1.65\pm0.33$\,mas and an inclination of $i\sim 0.5^\circ$, indicating
that the companion is a late type M-dwarf star with mass
$0.14\pm0.05$\msun.  In a large sample of ESP candidates, only a few
systems seen at low inclinations are expected.  It was therefore
surprising that, in a companion paper, \citet[hereafter HBG]{han01}
found, by combining {\it Hipparcos} data with the spectroscopic
orbital elements of 30 ESP candidates, that a significant fraction may
have small inclinations.  The semi-major axes of the orbits that HBG
derived are, however, at the limits of the precision of {\it
Hipparcos} measurements.  \citet{pourbaix01} and \citet{zucker01}
demonstrated that in these circumstances the small $i$'s were an
artifact of HBG's analysis. Indeed, \citet{mcgrath02} HST
astrometric observation of $\rho^1$ Cnc (55 Cnc), a star estimated by
HBG to have a particularly low inclination orbit, placed an upper
limit on its reflex motion, confirming that its companion is
sub-stellar.

We have been using high-resolution infrared (IR) spectroscopy to
detect low-mass companions in spectroscopic binaries first identified
in visible light (\citealt[hereafter M02]{mazeh02}; \citealt[hereafter
P02]{prato02}).  If \rcb's companion had the mass of a late M spectral
type main-sequence star, we expected it to lie within the sensitivity
limits of the technique \citep[e.g.][]{mazeh03}.  We describe here our
search for the companion in the \rcb{} system by IR spectroscopy. \S2
details extension of the IR technique to \rcb{}, our observations, and
data reduction. In \S3 we describe our analysis and the measurements
of \rcb.  In \S4 we set an upper bound on the 1.6\um{} flux of \rcb's
companion.  We calculate an upper limit on the companion mass using
theoretical models and comment on the significance of our result in
\S5.

\section{Observations and Data Reduction\label{data}}

We observed \rcb{} during four runs in the spring of 2001 (see
Table~1) with the W. M. Keck Observatory's cross-dispersed echelle IR
spectrometer NIRSPEC \citep{mclean98,mclean00} on the Keck II
telescope with and without adaptive optics (AO).  We used NIRSPEC in
its high resolution mode centered at 1.555\um.  This provided 9
spectral orders, 45~--~53, from 1.450~--~1.705\um{} with resolutions
of 28,000 and 35,000 in non-AO and AO modes respectively, for a 2
pixel slit width.  Further details of the instrument setup and
observation procedure are given in P02.  We extracted the spectra
using the
REDSPEC\footnote{http://www2.keck.hawaii.edu/inst/nirspec/redspec/index.html.}
software package, and determined the dispersion solution using OH
emission lines of the night sky, recorded simultaneously with the
observations and identified from the catalog of \citet{rousselot00},
and arc lamp spectra.

\begin{figure}
\plotone{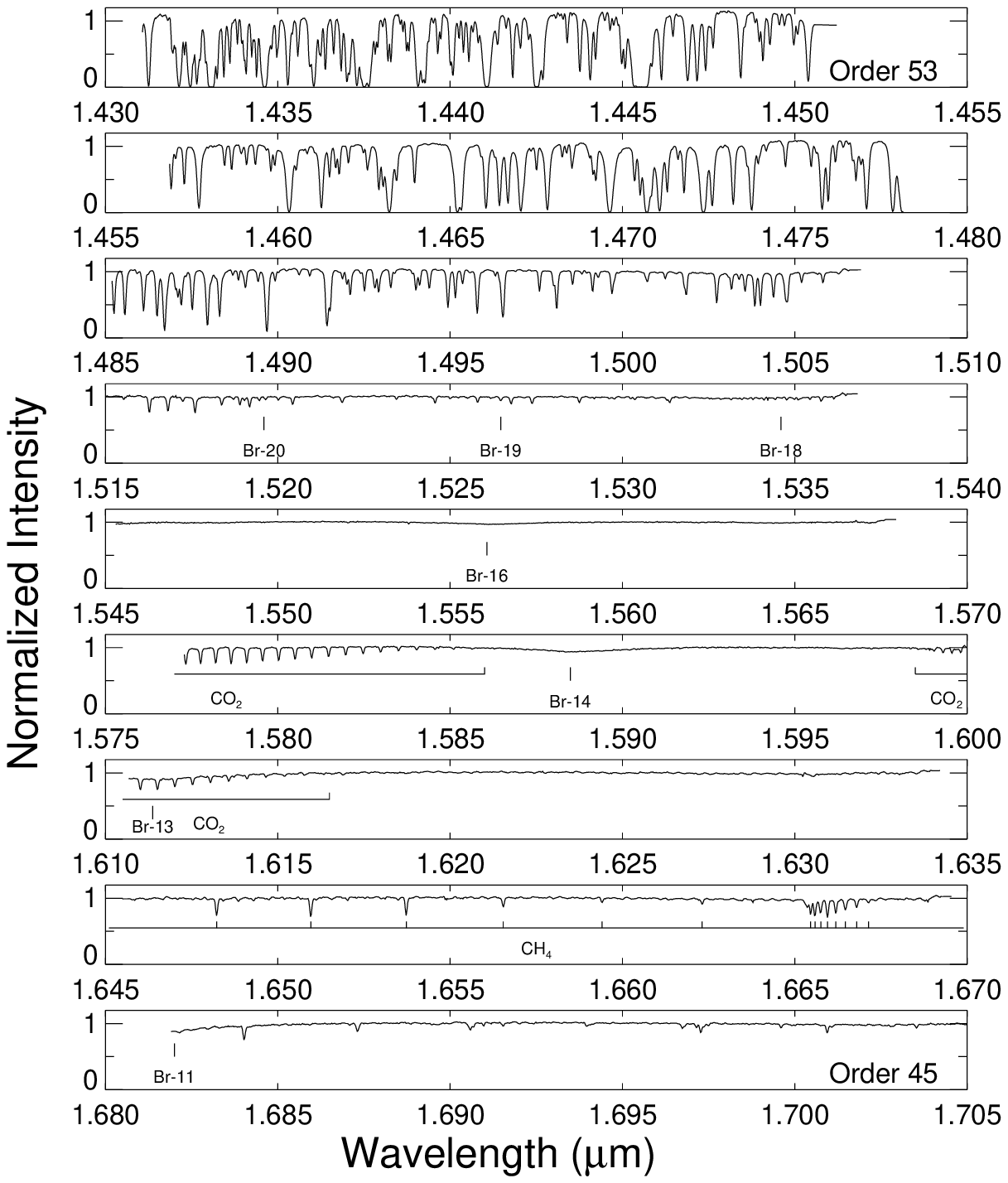}
\caption{NIRSPEC spectrum of the A0 star HD133772 for orders 45~--~53.
The spectra have been flattened and normalized to their continuum
levels.  The stellar spectra are nearly featureless except for the
very broad hydrogen Brackett lines indicated.  The remaining
absorption lines are terrestrial.  A $\mathrm{CH_4}$ band is indicated
in order 46 and two $\mathrm{CO_2}$ bands are marked in orders 47 and
48.\label{fig01}}
\end{figure}

Application of the IR spectroscopic technique to the detection of low
mass secondaries requires good templates for the primary and the
secondary.  Observations of these templates took place during 2000 and
2001 (P02).  Prior to this work (e.g., P02, M02), we used spectra in
only NIRSPEC order 49 because it is nearly completely free of
terrestrial absorption.  However, the sensitivity of our analysis
depends on the number of stellar spectral lines available and hence on
the spectral range of the data. To reach the low flux levels expected
for \rcb's companion we used additional spectral orders.
Figure~\ref{fig01} shows the spectrum of the A0 star HD 13372 in
orders 45~--~53.  The stellar spectrum contains only weak, broad lines
of the hydrogen Brackett series; all other absorption shown in the
figure is terrestrial.  The absence of terrestrial absorption in order
49 is apparent.  Order 46 contains terrestrial absorption in the P and
Q branches of a $\mathrm{CH_4}$ band; orders 47 and 48 contain
$\mathrm{CO_2}$ absorption lines.  Orders 51~--~53, and to a lesser
extent order 50, are contaminated by many terrestrial absorption
lines.  In order 45, with our choice of central wavelength, starlight
illuminated the array in only one of the two telescope nod positions.
We concentrated, therefore, on trying to correct atmospheric
absorption only in the spectra of orders 46, 47, and 48.

\begin{figure}
\plotone{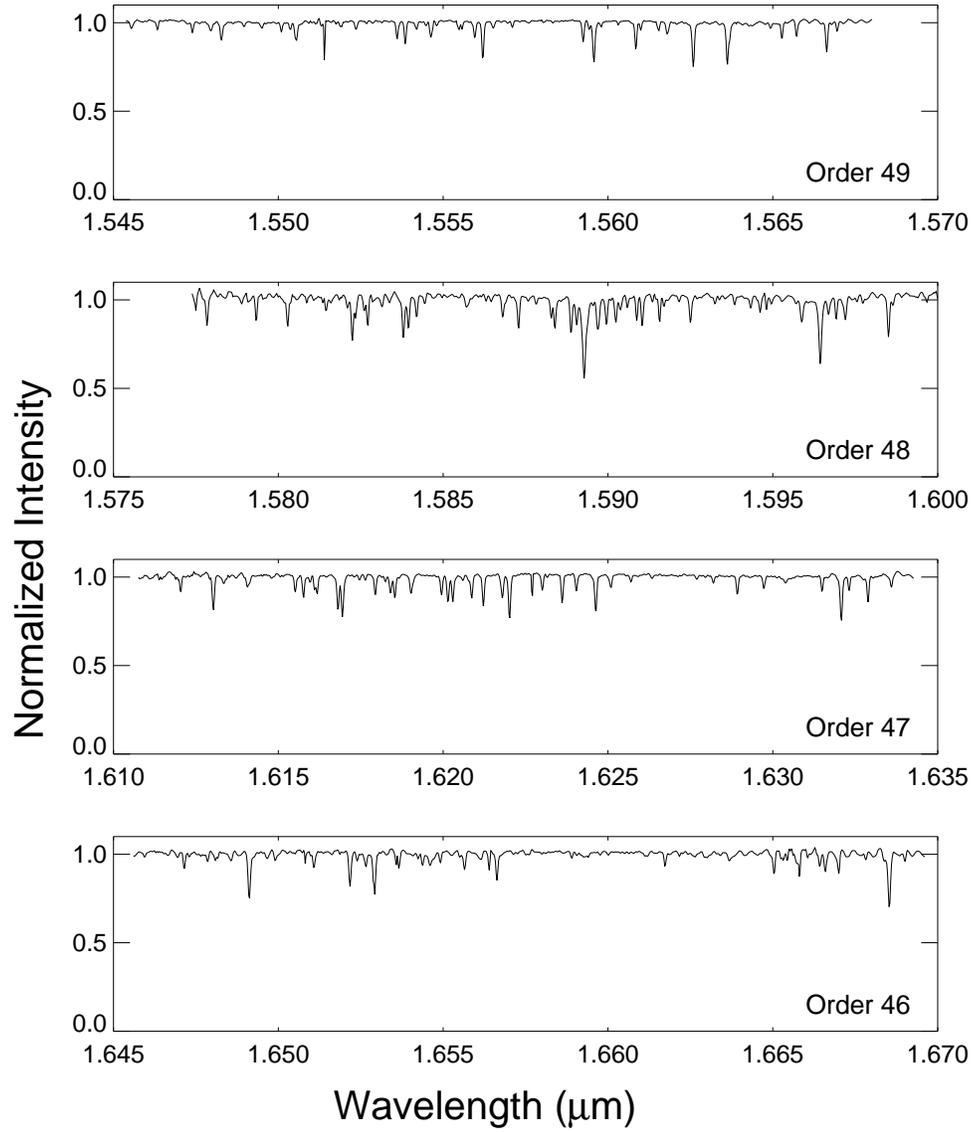}
\caption{NIRSPEC spectrum of \rcb{} from June 2, 2001 for orders
46~--~49.  The spectra have been flattened and normalized to their
continuum levels.  The terrestrial absorption lines in orders 46, 47,
and 48 have been removed by ratioing with calculated spectra (see
text).\label{fig02}}
\end{figure}

Terrestrial absorption lines are usually removed by dividing the
target spectrum by a featureless stellar spectrum observed nearly
simultaneously at the same airmass.  We took a different approach
because we had to correct many template spectra obtained in different
observing sessions.  We expect that the $\mathrm{CO_2}$ and
$\mathrm{CH_4}$ absorption bands are independent of weather and vary
only with airmass.  We therefore obtained NIRSPEC spectra of five A0
stars on July 17, 2002, over the range of airmasses of our \rcb{} and
template spectra. We removed the Brackett lines by fitting them with
Lorentzian profiles and used the terrestrial absorption lines in
orders 46, 47, and 48 to register each observation on a sub-pixel
scale.  For each of the terrestrial lines, we fitted the dependence of
the absorption on airmass, and were thus able to calculate terrestrial
absorption spectra at the exact airmasses of the \rcb{} and template
spectra.  This procedure worked well for order 47 and less well in
orders 46 and 48, probably because their telluric lines are deeper and
have more complex structure than those in order 47.  The residuals in
orders 46 and 48 appear in only a few percent or less of free spectral
range of these orders so we considered them negligible.
Figure~\ref{fig02} shows the order 46~--~49 \rcb{} spectra measured on
June 2, 2001, with orders 46, 47, and 48 corrected for telluric
absorption.

\begin{figure}
\plotone{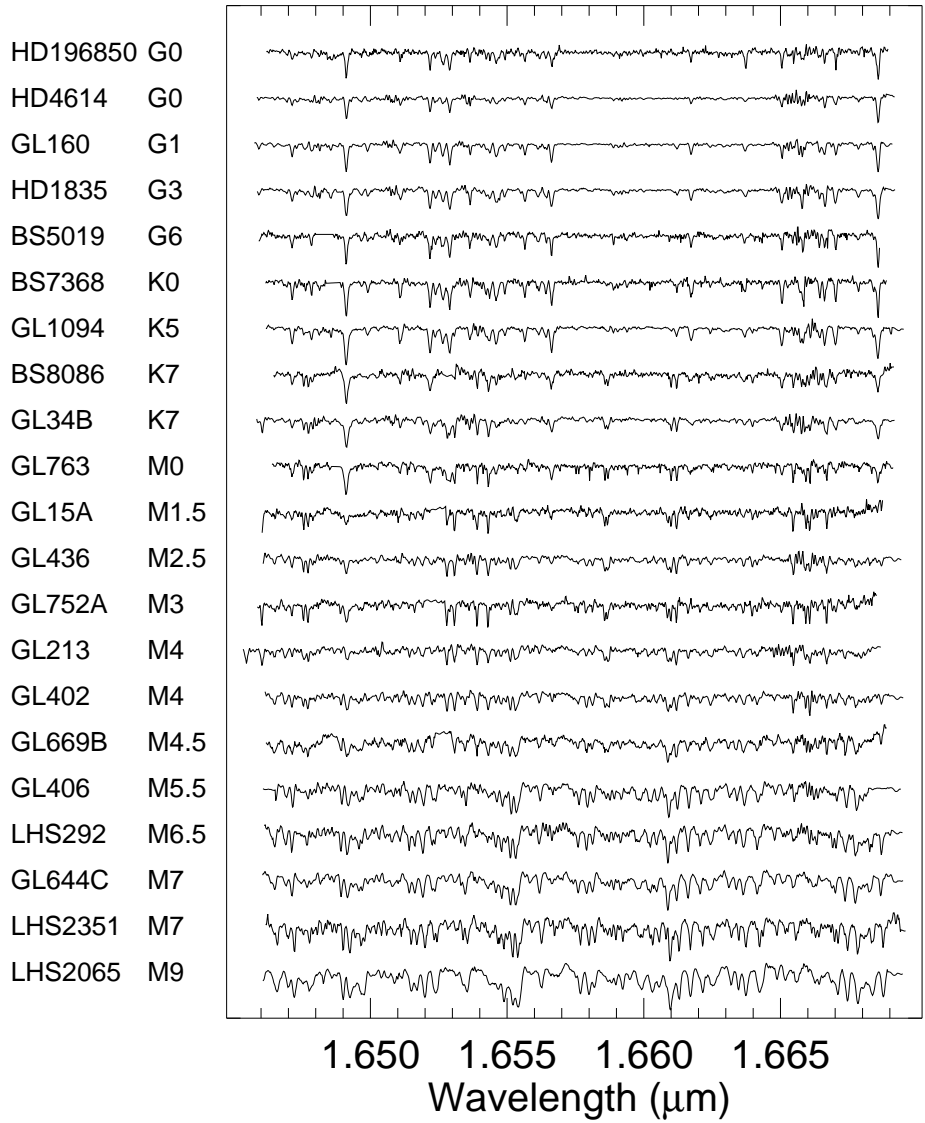}
\caption{NIRSPEC order 46 templates.  The spectra have been flattened,
  normalized to their continuum levels, and terrestrial absorption
  removed.\label{fig03}}
\end{figure}

\begin{figure}
\plotone{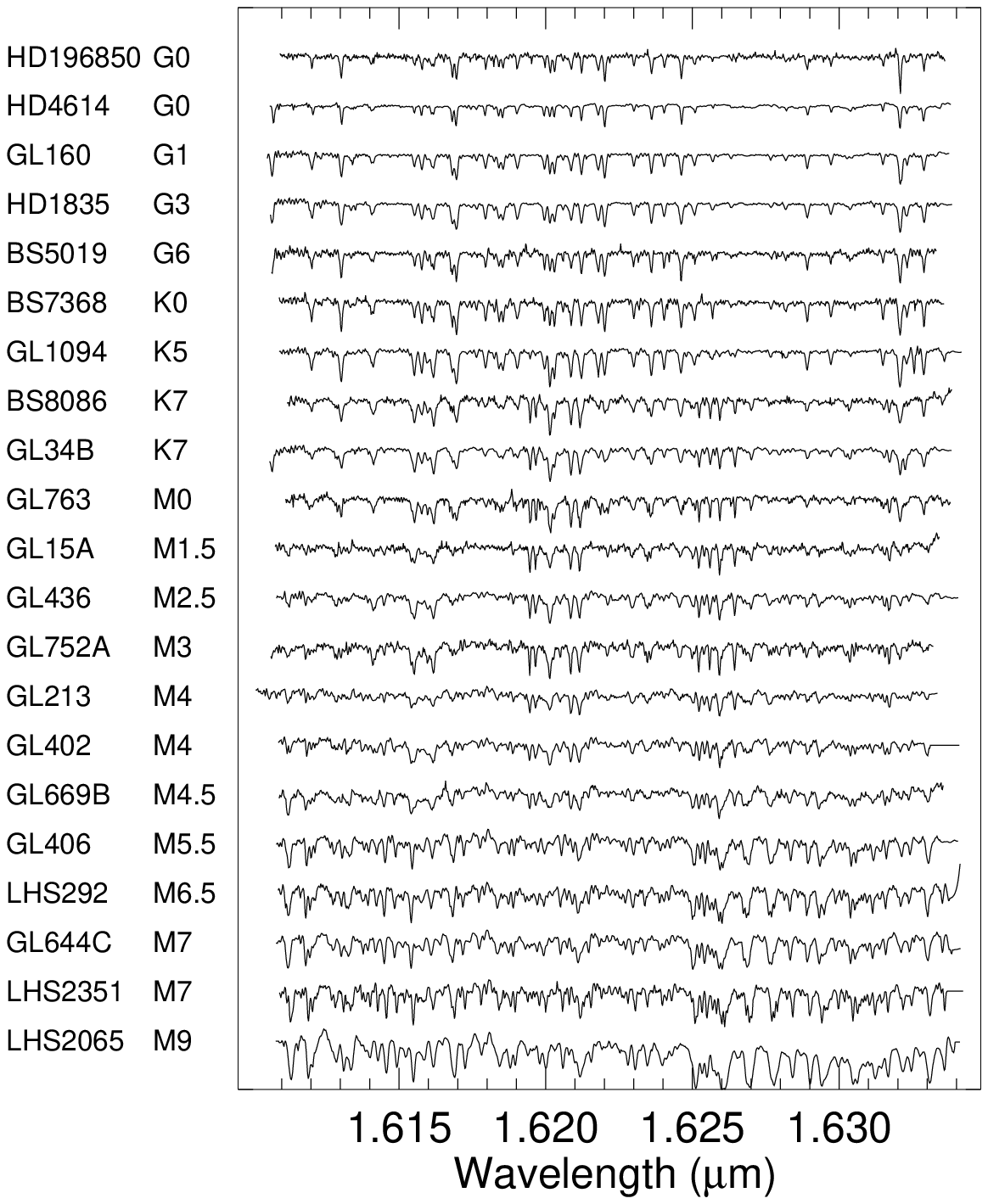}
\caption{Same as Figure~\ref{fig03}, but for NIRSPEC order
  47.\label{fig04}}
\end{figure}

\begin{figure}
\plotone{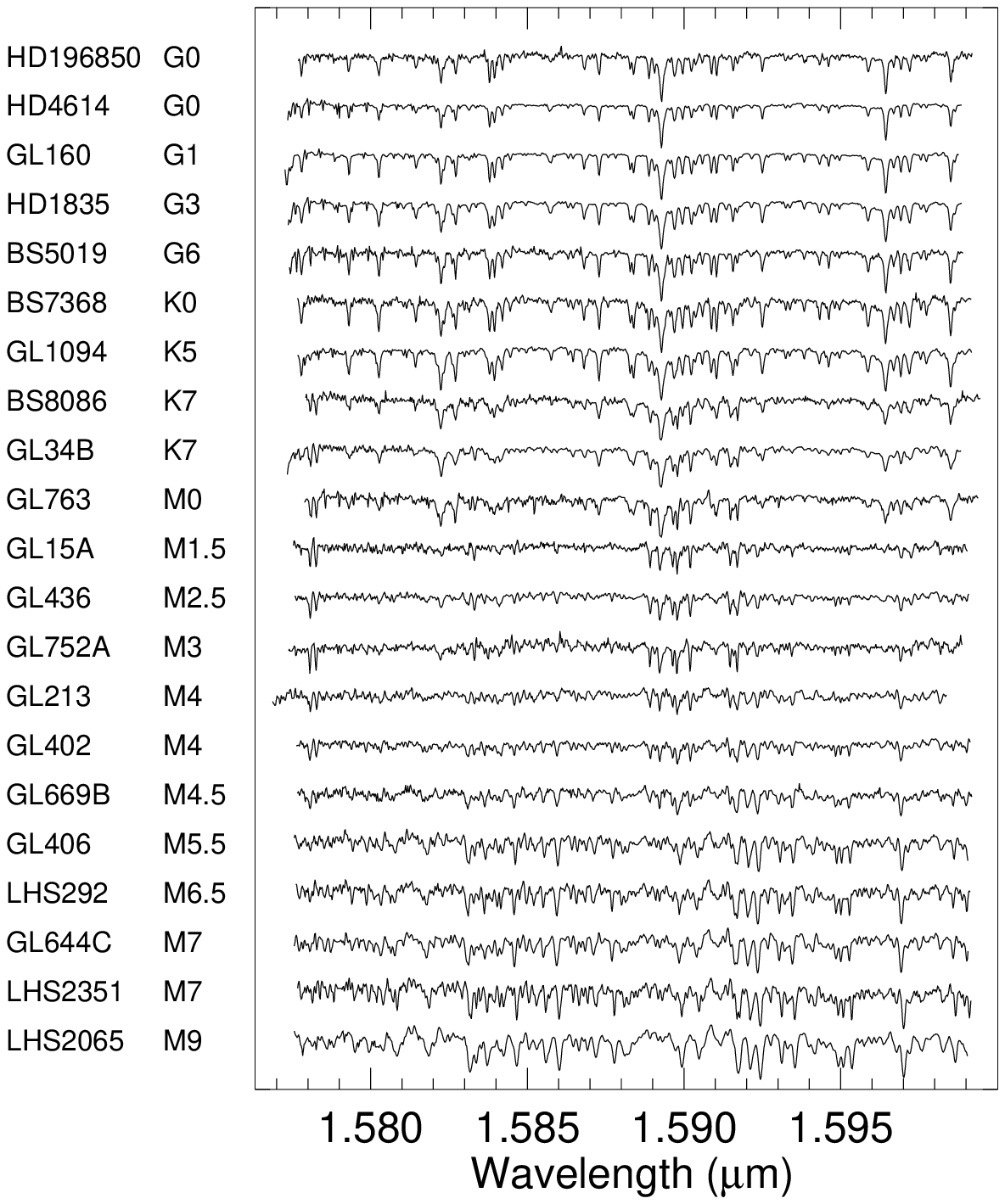}
\caption{Same as Figure~\ref{fig03}, but for NIRSPEC order
  48.\label{fig05}}
\end{figure}

\begin{figure}
\plotone{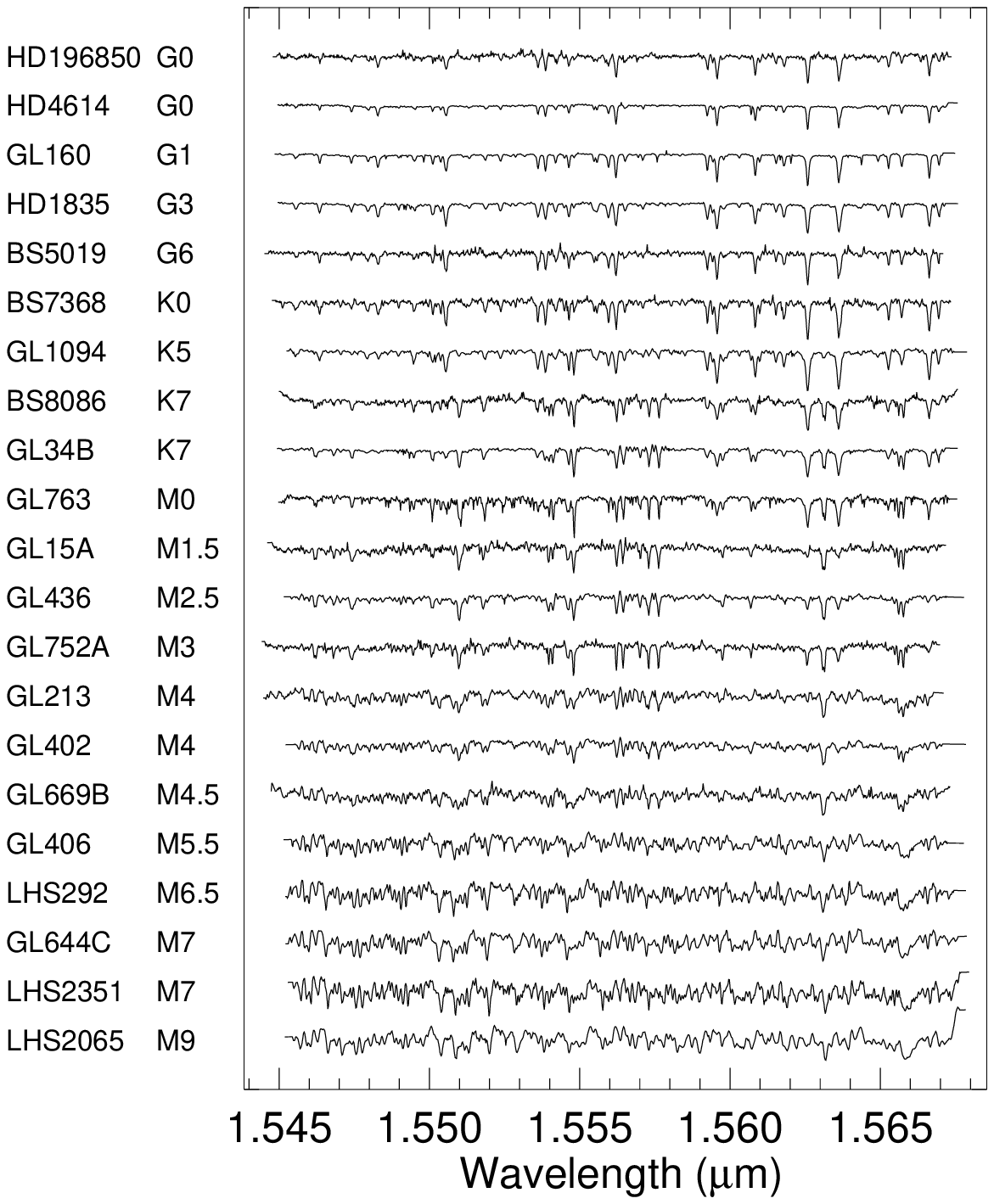}
\caption{Same as Figure~\ref{fig03}, but for NIRSPEC order 49.
\label{fig06}}
\end{figure}

To ensure consistent results, we reextracted orders 46~--~49 for the
entire template set, calculated the dispersion solution using the OH
lines of the night sky, corrected the spectra for terrestrial
absorption, and rederived the stellar radial velocities following the
technique described in P02. Figures~\ref{fig03}~--~\ref{fig06} show
the template spectra for orders 46~--~49 in the laboratory reference
frame, with the terrestrial absorption lines removed from orders 46,
47, and 48.

\section{The Companion/Primary Flux Ratio of \rcb\label{analysis}}

Our procedure to search for evidence of \rcb's companion is similar to
that described by M02 for spectroscopic binaries using the
two-dimensional cross-correlation routine TODCOR \citep{zucker94}.
The analysis identifies the templates that provide the best matching
primary and secondary, the ratio of the secondary and primary fluxes,
$\alpha$, and the radial velocities of the components.  The measured
amplitude of \rcb's reflex velocity is 67\mps{} (N97).  If its
companion had the 0.14\msun{} mass reported by GHB, the maximum
velocity difference of the primary and secondary would be
0.4\kps. This is less than the 1\kps{} velocity precision of our
technique (P02), so we did not expect to detect a velocity difference.
To obtain a velocity difference large enough for us to detect
reliably, $>1$\kps, would require that the companion have mass
$\lesssim 0.05$\msun; the flux of such a substellar object blended
with a G star would be too small for us to detect spectroscopically.
We therefore regarded the velocity difference of the components as
fixed at 0 and focused on measuring the system's flux ratio.

We wrote a routine to model binaries from pairs of templates at
prescribed flux ratios, to cross-correlate them with our \rcb{}
spectra, and to identify the model most closely matching \rcb.  With
this, we analyzed separately the \rcb{} spectra in orders 46~--~49 for
each of the four observing sessions using a wide selection of
templates. The highest correlation values resulted from using HD~4614
or GL~160 for the primary template, depending on the particular \rcb{}
spectrum, and the late M-type templates for the secondary.  Changing
the secondary spectral type by a few subclasses resulted in negligible
changes in the correlation, and we concluded that we could only
discriminate between early and late type M-star secondaries.  Based on
this, we averaged the $\alpha$'s resulting from the late type
secondaries GL~406, LHS~292, GL~644C, LHS~2351, and LHS~2065 for each
\rcb{} spectrum.  These averaged values are our measured flux ratios
and are listed in Table~2.  The arithmetic mean of the measured flux
ratios in Table~2 is $-$0.0034.  This small value, and the large
scatter around it, from $-$0.0260 to 0.0164, suggest that the
observations have not detected the companion.  The primary and
secondary templates are not likely to match exactly the effective
temperature and metallicity of the \rcb{} primary and companion.  The
fact that the average flux ratio is essentially zero indicates that
any such mismatch is not causing a systematic error.

To interpret the apparent non-detection we must derive an upper bound,
at a specified level of significance, on the flux ratio.  The
conventional approach would be to calculate the mean of the measured
flux ratios in Table~2, to assume they follow a Gaussian distribution
and derive the standard deviation of the mean, and to use this to
place confidence limits on an upper bound.  The measured flux ratios,
however, do not necessarily follow a Gaussian distribution, and the
number of measurements, 16, is too few to test the assumption
reliably.  We chose, therefore, to take a different approach, using
the data itself to model a large number of observations, and from
these directly measure the probabilities of a detection at a certain
flux ratio, independent of underlying statistical assumptions.  \S4
describes this approach.

\section{Estimate of the Flux Ratio Upper Bound\label{upperlimit}}

\subsection{Model Binaries\label{models}}

\begin{figure}
\plotone{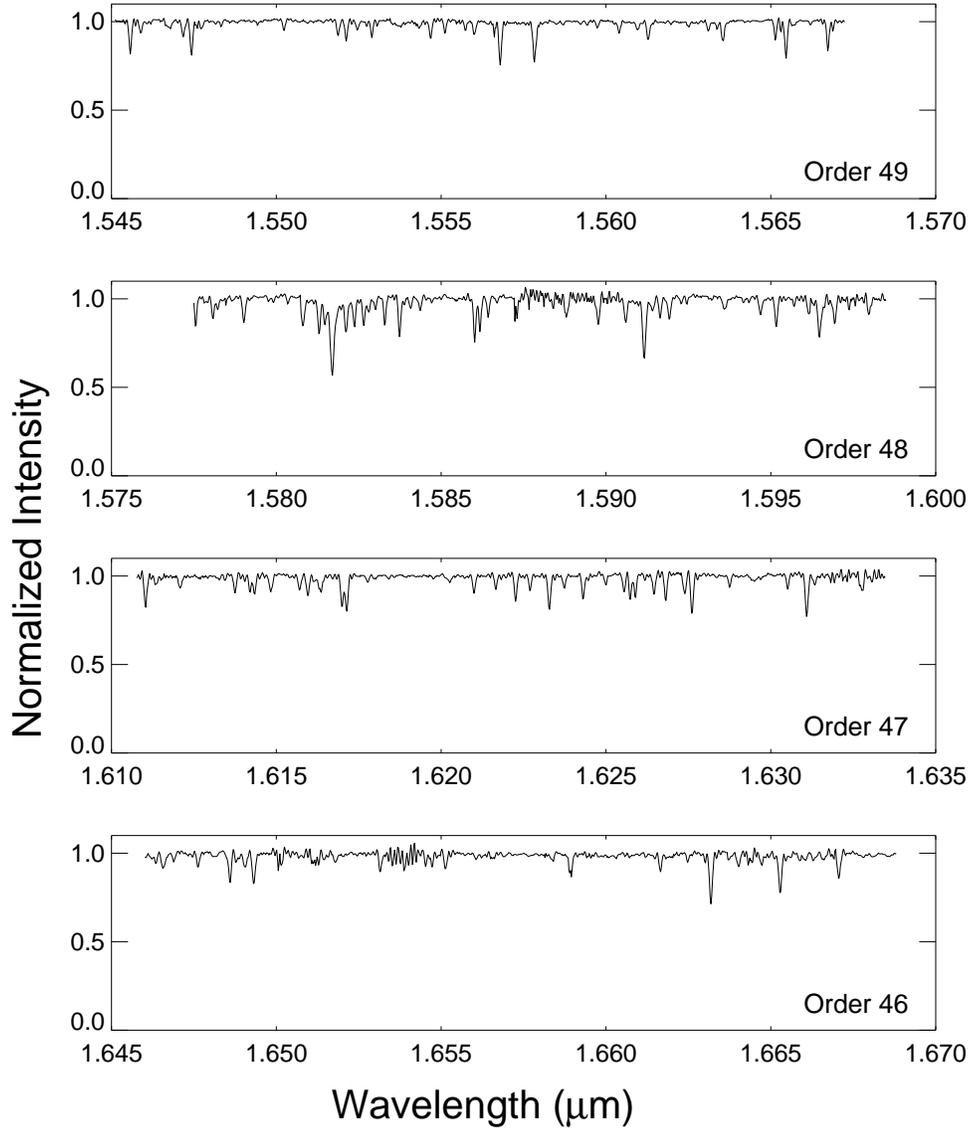}
\caption{Example model binary primary spectra from the January 2001
model set.  The \rcb{} spectra were subdivided, at wavelengths with no
spectral lines, into smaller sections with lengths 10~--~60\,\AA.  The
sections were reassembled randomly to create unique spectra with
absorption lines, noise characteristics, and spectral range identical
to the original \rcb{} spectra.\label{fig07}}
\end{figure}

In this section, we estimate the upper bound on the ratio of the
companion's flux to that of \rcb ~that follows from the measurements
in Table~2.  Our approach starts by analyzing sets of model binaries.
To ensure the models closely represented the original observations, we
used the observed \rcb{} spectra as the basis for the model primaries.
We subdivided each observed \rcb{} spectrum at wavelengths where there
are no lines, into smaller sections with lengths from 10 to 60\,\AA.
We reassembled these sections in random sequences to create a set of
500 unique model primaries for each observation, with absorption
lines, noise characteristics, and free spectral range identical to the
original \rcb{} spectrum.  Figure~\ref{fig07} shows examples of
reassembled model primaries derived from the January 2001 \rcb{}
spectra. We used the M7 spectral type star LHS~2351 for the secondary
in all of the models, and combined it, unmodified, with the randomized
primaries at eight flux ratio values, from 0.001~--~0.020, to create
the model binary spectra.  This provided us with 128 sets (4
observations $\times$ 4 orders $\times$ 8 flux ratios) of 500 model
binaries each, with which to evaluate the measured flux ratios in
Table~2.

We analyzed the model binaries in the same way as the original \rcb{}
spectra (\S\ref{analysis}).  For each set of models, we cut and
reassembled a matching set of primary templates from either HD~4614 or
GL~160, depending on which was used in the corresponding \rcb{}
analysis.  We restricted the analysis to only the M6.5 spectral type
star LHS~292 for the secondary template because the correlation value
is insensitive to small changes in the secondary spectral type
(\S\ref{analysis}).

\begin{figure}
\plotone{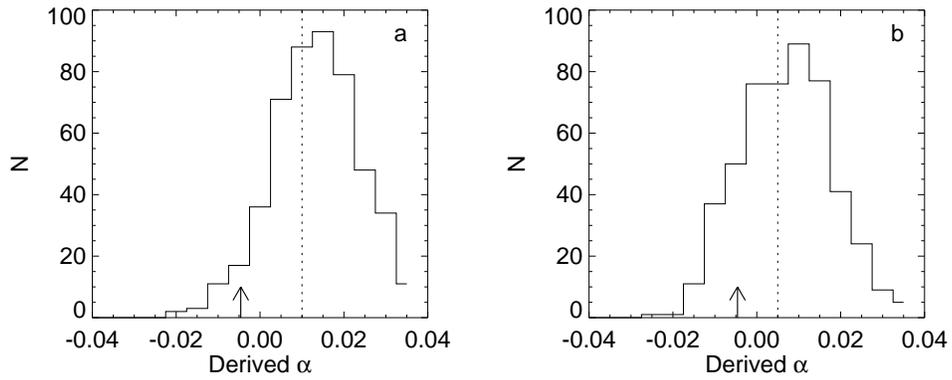}
\caption{{\it a:} The distribution of derived flux ratios in a sample
of 500 model binaries having the characteristics of the \rcb{} order
49 spectrum measured in January 2001 and the companion GL 406 at flux
ratio $\alpha=0.010$.  The dashed line indicates the input $\alpha$
and the arrow the measured value, $-0.0046$ (Table~2).  {\it b:} Same as
{\it(a)} but for input $\alpha=0.005$. \label{fig08}}
\end{figure}

The analysis produced 500 ``measured'' flux ratios for each of the 128
sets of models.  For a given set of models, the measured flux ratios
scattered around the input value.  Figure~\ref{fig08} shows two of these
distributions from January 2001, order 49, $\alpha=0.010,0.005$, along
with the corresponding measured value of $-$0.0046 from Table~2.

\subsection{Flux Ratio Upper Bound\label{fisher}}

From each of our model distributions we can determine the probability
that the corresponding \rcb{} observation returns a flux ratio as
small as the one measured.  Consider, for example, the results for
January 2001, order 49.  If the actual flux ratio were 0.010, the
probability that the observation and its analysis produces a flux
ratio as small as the one measured, $-$0.0046, is given by the number
of model binaries that returned flux ratios of equal value or smaller.
Figure~8a shows the model distribution derived for this case; 24 out
of 500 models (probability=0.048) produced an $\alpha$ smaller than
$-$0.0046.  If, then, the true $\alpha$ were 0.010, the probability of
a measurement producing a value as low as or lower than $-$0.0046 is
unlikely, and we could consider 0.010 an upper bound with 95\%
confidence $(1-0.048\sim0.95)$.  Suppose that instead the true
$\alpha$ were smaller, 0.005.  In this case, a larger number of
models, now 76 out of 500, or 15\%, give an $\alpha$ as small as the
measured one (Figure~8b).  We would therefore consider 0.005 as an
upper bound with only 85\% confidence.

\begin{figure}
\plotone{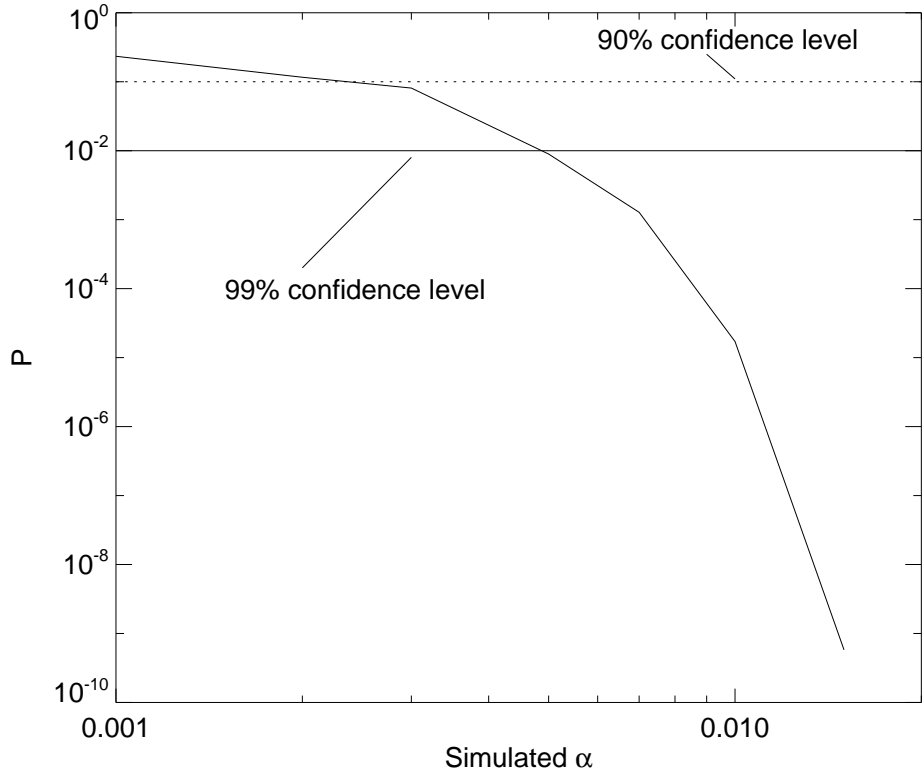}
\caption{The curve shows the combined probabilities that our 16
measurements (4 spectral orders on 4 occasions) have a flux ratio less
than or equal to the measured value, as a function of the model binary
input flux ratio. The straight lines with constant probability show
the 90\% and 99\% confidence levels and intersect the curve of combined
probabilities at a flux ratios of 0.0024 and 0.005.\label{fig09}}
\end{figure}

For each of the 16 modeled observations, we calculated, at the eight
input model flux ratios, the probability of measuring a flux ratio as
low as or lower than the measured value in Table~2.  To combine those
probability values into one value, we use Fisher's method to combine
independent $p$-values \citep{fisher32}.  In this method, one first
calculates the product of probabilities.  This value, according to
\citet{fisher32}, is distributed in a known way, and we can use its known
distribution to infer the new probability.  The product value, $k$, is
\begin{equation}k=\prod_{i=0}^{n-1}p_i.\end{equation} \citet{fisher32}
showed that the statistic $F=-2\ln{k}$ has a $\chi^2$ distribution
with $2n$ degrees of freedom.  Following Fisher's recipe\footnote{See
also http://www.loujost.com}, we use this known distribution to
calculate $P$, the combined $p$-value from the 16 probabilities.
Figure~\ref{fig09} shows $P$ as a function of input flux ratio.  The
combined probability curve intersects the 99\% confidence level at
$\alpha=0.005$.  Similarly, the flux ratio 0.0024 can be ruled out
with 90\% confidence.

\section{Discussion}

We consider the flux ratio from 1.55 to 1.65\um{} as equivalent to
that in the $H$-band because spectra of late M dwarfs between 1.4 and
1.7\um{} are not marked by the very deep H$_2$O absorption
characteristic of L spectral type dwarfs \citep[e.g.][]{leggett01}.
The 2MASS $H$-band magnitude of \rcb{} is 3.99.  This is in excellent
agreement with its {\it Hipparcos} V magnitude, 5.39 (HIP 78459), and
the $\mathrm{V-H}$ color, 1.36, for a GOV spectral type star
\citep{tokunaga00}.  With the {\it Hipparcos} distance of
$17.3\pm0.2$\,pc, \rcb{}'s absolute H-band magnitude is
$\mathrm{M_H=2.8}$.  The flux ratio upper bound (\S4.2) therefore
corresponds to an $H$-band brightness limit for the companion of
$\mathrm{M_H=8.6}$ at the 99\% confidence level and $\mathrm{M_H=9.3}$
at the 90\% confidence level.

\begin{figure}
\plotone{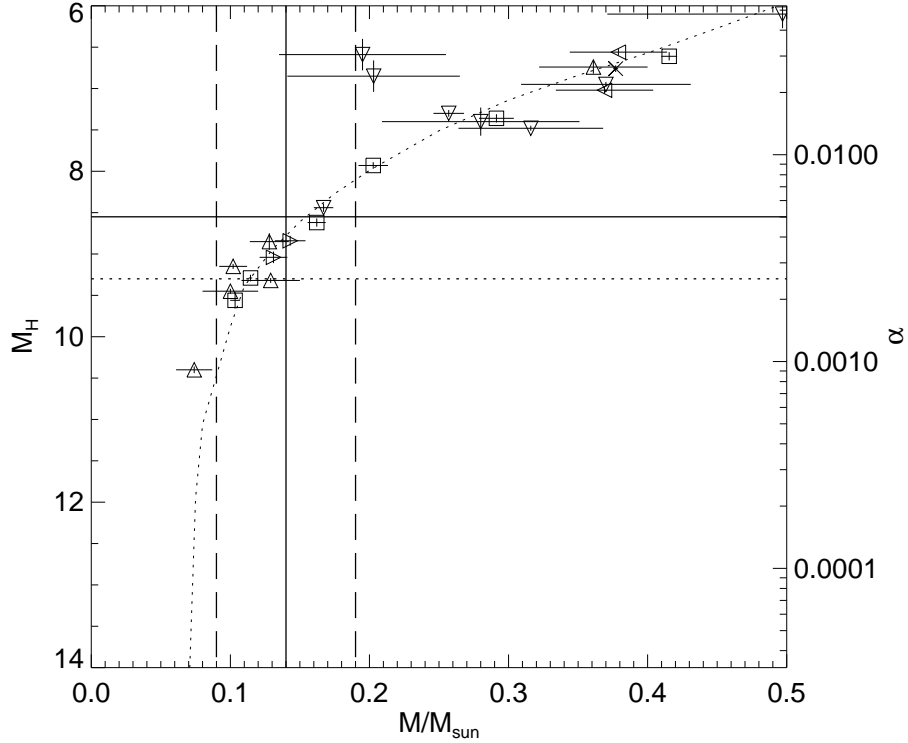}
\caption{Absolute H-band magnitude vs. mass; the (companion/primary)
flux ratio at 1.6\um, $\alpha$, is also shown for reference.  The
dotted curve corresponds to BCAH98's 10\,Gyr isochrone, with values
for M/\msun$\leq0.1$ updated by Baraffe et al. (2003).  The vertical
lines with constant mass represent GHB's result and $1\sigma$ error
estimate.  The horizontal lines at $\mathrm{M_H=8.6}$ and 9.3
represent our flux ratio limits of 0.005 and 0.0024 for \rcb's
companion at the 99\% and 90\% confidence limits, and corresponds to a
mass upper bounds of 0.15 and 0.11\msun, respectively.  The data
points are compiled from the literature and the symbols correspond to
their source as follows: $\vartriangle$ -- \citet{henry99};
$\triangledown$ -- \citet{henry93}; $\square$ -- \citet{segransan00};
$\times$ -- \citet{forveille99}; $\triangleright$ -- \citet{torres99};
$\triangleleft$ -- \citet{martin98}.  $\mathrm{M_H}$ values come from
these and \citet{delfosse00}, \citet{leggett02}, and
\citet{leggett92}.\label{fig10}}
\end{figure}

Figure~\ref{fig10} shows the absolute H magnitude versus mass, 10\,Gyr
isochrone calculated by \citet{baraffe98} for low mass stars and brown
dwarfs, with the model values for masses less than 0.1\msun{} updated
from \citet{baraffe03}.  At the scale of the figure, the 5\,Gyr
isochrone is indistinguishable from the one for 10\,Gyr. We also plot
the locations of main-sequence dwarfs whose masses have been measured
dynamically; the references are cited in the figure caption.  For
reference, the figure shows the (companion/primary) flux ratio upper
bounds calculated in \S\ref{upperlimit}.  The flux ratio upper bound
at the 99\% confidence level, $\alpha=0.005$, falls at
$\mathrm{M\sim0.15}$\msun, close to GHB's reported mass of the
companion, $0.14\pm0.05$\msun.  Similarly, $\alpha=0.0024$ sets an
upper bound on the mass of $M\sim0.11$\msun{} with 90\% confidence.
While these mass upper bounds do not provide either a definitive
confirmation or rejection of GHB's reported mass of the companion,
they do suggest strongly that its mass is smaller than their value.

The analysis described in \S3 and \S4, and illustrated by the
histograms in Figure~\ref{fig08} shows that, by increasing the
spectral range through the use of several orders, the IR spectroscopic
technique can detect binaries with 1.6\um{} flux ratios in the range
of 0.01 to 0.02.  This is a significant advance over our previous
detections at flux ratios of 0.04 to 0.05 (Mazeh et al. 2003).

\section{Summary\label{Summary}}

1) Our measurements set upper limits of 0.0024 and 0.005 on the
ratio of the $1.6$\um{} flux of \rcb's companion to \rcb, at the
$90\%$ and $99\%$ confidence levels, respectively.

2) Using Baraffe et al.'s (1998, 2003) calculations of the $H$-band
mass-luminosity relation, these flux limits correspond to mass upper
bounds of 0.11 and 0.15\msun{} at the 90 and 99\% confidence limits,
respectively.

3) Our analysis of the model binaries tested the sensitivity of our
technique when using multiple orders.  The results indicate that
binaries with 1.6\um{} flux ratios as low as 0.01 to 0.02 are
detectable by high spectral resolution IR spectroscopy.

\vskip 1.0cm

We thank the referee for useful comments that improved the
manuscript. MS thanks D. Black for the chance conversation that
started this project.  We thank the staff of the W.M. Keck Observatory
for their thorough support and C. Koehn for help with extraction of
the template spectra.  The work of CB and MS was supported in part by
NSF grant AST 02-05427.  TM was supported in part by the Israeli
Science Foundation grant 233/03.  SZ is grateful for partial support
from the Jacob and Riva Damm Foundation.  Data presented herein were
obtained at the W.M. Keck Observatory which is operated as a
scientific partnership between the California Institute of Technology,
the University of California, and NASA.  The Observatory was made
possible by the generous financial support of the W.M. Keck
Foundation.  We extend our thanks to those of Hawaiian ancestry on
whose sacred mountain we are privileged to be guests.  This
publication makes use of data products from the Two Micron All Sky
Survey, which is a joint project of the University of Massachusetts
and the Infrared Processing and Analysis Center/California Institute
of Technology, funded by the National Aeronautics and Space
Administration and the National Science Foundation.  This research has
also made use of the SIMBAD database, operated at CDS, Strasbourg,
France.

\clearpage

\epsscale{0.8}

\clearpage

\begin{deluxetable}{lcc}
\tablecaption{Summary of \rcb{} Observations \label{tblobs}}
\tablewidth{0pt}
\tablehead{
\colhead{UT Date} &
\colhead{Obs Mode} &
\colhead{Total Exp. Time (s)}}
\startdata
2001 Jan 7 & AO & 640 \\
2001 Feb 2 & non-AO & 960 \\
2001 May 3 & non-AO & 840 \\
2001 Jun 2 & non-AO & 840 \\
\enddata
\end{deluxetable}

\clearpage

\begin{deluxetable}{lcr}
\tablecaption{Companion/Primary Flux Ratios at 1.6\um \label{tblresults}}
\tablewidth{0pt}
\tablehead{
\colhead{Observation} & \colhead{NIRSPEC} &  \\
\colhead{Date} & \colhead{Order} & \colhead{$\alpha$}}
\startdata
2001 Jan 7 & 46 & $-0.0100$ \\
           & 47 & $ 0.0125$ \\
           & 48 & $-0.0090$ \\
           & 49 & $-0.0046$ \\
2001 Feb 2 & 46 & $-0.0084$ \\
           & 47 & $ 0.0068$ \\
           & 48 & $-0.0114$ \\
           & 49 & $ 0.0164$ \\
2001 May 3 & 46 & $-0.0148$ \\
           & 47 & $ 0.0065$ \\
           & 48 & $-0.0148$ \\
           & 49 & $-0.0012$ \\
2001 Jun 2 & 46 & $-0.0260$ \\
           & 47 & $ 0.0118$ \\
           & 48 & $-0.0020$ \\
           & 49 & $-0.0054$ \\
\enddata
\end{deluxetable}

\end{document}